\documentclass[aps,reprint,prb,twocolumn,superscriptaddress,floatfix]{revtex4-2}%
\usepackage{graphicx}
\usepackage{color}
\usepackage{lipsum}
\usepackage{braket}
\usepackage{amsmath}
\usepackage{pifont}
\usepackage[colorlinks=true,citecolor=green,linkcolor=blue]{hyperref}
\usepackage{color}

\newcommand{\bt}{\textbf}

\begin{document}

\title{Quantum transport in an ambipolar InSb nanowire quantum dot device}
\author{Mingtang~Deng}
\email{mtdeng@nudt.edu.cn}
\affiliation{College of Computer Science and Technology, NUDT, Changsha 410073, China}
\affiliation{Division of Solid State Physics, Lund University, Box 118, S-221 00 Lund, Sweden}
\affiliation{Hefei National Laboratory, Hefei 230088, China}
\author{Chunlin~Yu}
\affiliation{China Greatwall Quantum Laboratory, Changsha 410006, China}
\affiliation{Division of Solid State Physics, Lund University, Box 118, S-221 00 Lund, Sweden}
\author{Guangyao~Huang}
\affiliation{College of Computer Science and Technology, NUDT, Changsha 410073, China}
\author{Philippe~Caroff}
\affiliation{Division of Solid State Physics, Lund University, Box 118, S-221 00 Lund, Sweden}
\author{Hongqi~Xu}
\email{hqxu@pku.edu.cn}
\affiliation{Division of Solid State Physics, Lund University, Box 118, S-221 00 Lund, Sweden}
\affiliation{Beijing Key Laboratory of Quantum Devices, Key Laboratory for the Physics and Chemistry of Nanodevices, and Department of Electronics, Peking University, Beijing 100871, China}
\affiliation{Beijing Academy of Quantum Information Sciences, Beijing 100193, China} 

\begin{abstract}
Semiconductor InSb nanowires present a highly intriguing platform with immense potential for applications in spintronics and topological quantum devices. The narrow band gap exhibited by InSb allows for precise tuning of these nanowires, facilitating smooth transitions between the electron transport region and the hole transport region. In this study, we demonstrate quantum transport measurements obtained from a high-quality InSb nanowire quantum dot device. By utilizing a back gate, this device can be adjusted from an electron-populated quantum dot regime to a hole-populated one. Within both regimes, we have observed dozens of consecutive quantum levels without any charge rearrangement or impurity-induced interruptions. Our investigations in the electron transport regime have explored phenomena such as Coulomb blockade effect, Zeeman effect, and Kondo effect. Meanwhile, in the hole-transport regime, we have identified conductance peaks induced by lead states. Particularly, we have created a tomographic analysis method of these lead states by tracking the behavior of these conductance peaks across consecutive Coulomb diamond structures.
\end{abstract}

\date{\today}

\maketitle
Single-electron and single-hole semiconductor quantum dot devices have emerged as promising candidates for qubit realization due to their controllable two-level spin states (spin-up and spin-down), as evidenced in numerous studies~\cite{Loss1998, Petta2005, Nowack2007, Brunner2009, Hu2012, Pribiag2013}. Transport measurements on electron and hole quantum dots provide valuable insights into the potential of these systems for implementing quantum computing. Realizing both single-electron and single-hole quantum dots within a single device through gating offers significant flexibility and innovation. However, integrating both $n$-type and $p$-type quantum dots into a single quantum circuit presents substantial challenges. The main difficulty lies in fabricating high-quality quantum devices that can accommodate both carrier types, as materialization and fabrication processes are often optimized specifically for one carrier type during doping or work-function engineering procedures~\cite{Zheng2004, Jin2004, Wang2011, Heinze2002, Giovannetti2008, Leonard2011}.

Meanwhile, the utilization of narrow band-gap materials for engineering ambipolar devices is feasible. III-V compound semiconductors with narrow band gaps, particularly InAs and InSb, offer promising platforms that can accommodate both $n$- and $p$-type polarities~\cite{Doornbos2010, Lei2023}. Due to their exceptional properties such as small effective carrier masses, high carrier mobilities, large $g$-factors, and pronounced spin-orbit interactions, InAs and InSb have also garnered significant attention as potential platforms for topological quantum computing~\cite{Mourik2012, Deng2012, Das2012, Deng2014, Albrecht2016, Deng2016, Deng2018, Fu2021}. The realization of ambipolar InAs or InSb devices can be achieved through gate tuning, further highlighting their versatility and potential in future quantum technologies.

In this study, we have fabricated and characterized an InSb nanowire quantum dot device integrated with Nb-based contacts. By exploiting the narrow band-gap of the InSb nanowire, we are able to precisely control the quantum dot's occupation from the conductance-band-electron populated regime (the electron-transport or $n$-type regime) to the valence-band-hole populated regime (the hole-transport or $p$-type regime). Within the $n$-type regime, we observe up to 50 consecutive quantum levels exhibiting a remarkably regular alternating pattern of large and small level spacings. Conversely, within the $p$-type regime, we detect up to 30 continuously resonant quantum levels with nearly uniform additional energies. The exceptional quality of our material growth and meticulous fabrication process ensures exceptionally clean charge spectra in both regimes without any impurity-induced disruptions within our measured range. Furthermore, our investigations delve into Zeeman splitting, Kondo-effect correlations, and lead-state-induced conductance peaks within the Coulomb blockade spectra.

The Nb-InSb nanowire quantum dot-Nb device investigated in this study [Fig.~\ref{Fig1}(a)] has also been reported in the supplementary material of Ref.~\onlinecite{Deng2014}, where the superconductivity-related physics in the $n$-type region is discussed. To distinguish this device from the one described in the Appendix of the present article, we refer to the device in Fig.~\ref{Fig1}(a) as Device A. Device-A is fabricated from a zincblende InSb nanowire segment of an epitaxially grown InSb/InAs heterostructure nanowire. The nanowire is deposited onto a heavily doped Si/SiO\textsubscript{2} substrate, which features a pre-fabricated Ti/Au (5/100 nm) layer at the back side serving as a global gate electrode. Ti/Nb/Al (3/80/5 nm) contacts with a spacing of approximately 100 nm are sputtered on both ends of the InSb nanowire segment. The 3-nm Ti is a sticking layer, while the 5-nm Al layer protects the Nb from oxidation. It should be noted that prior to sputtering the Ti/Nb/Al contacts, the nanowire undergoes treatment in an ammonia sulfide solution to eliminate the native oxide layer and passivate the fresh semiconductor surface. Subsequently, measurements are conducted on the device within a wet $^3$He/$^4$He dilution refrigerator with a base temperature of 25 mK. For more comprehensive information regarding material growth, device fabrication, and measurement setup, please refer to Ref.~\onlinecite{Deng2014} and its associated references.

\begin{figure}
\centering
\includegraphics[width=8.5cm]{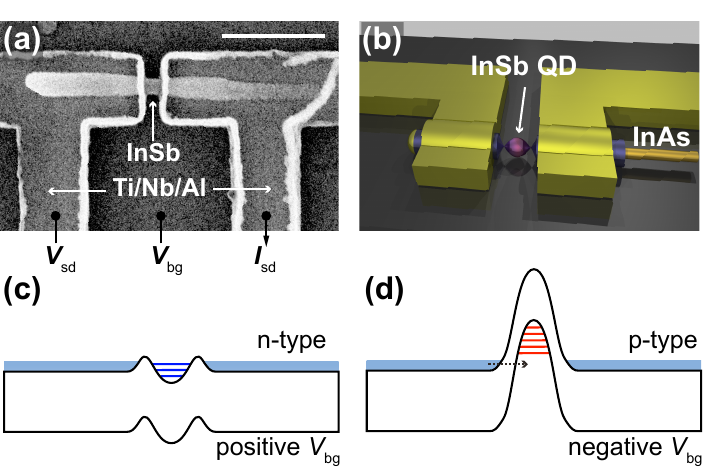}
\caption{\label{Fig1} \bt{Device overview}. (\textbf{a}) Scanning electron microscopy image and (\textbf{b}) schematic diagram of the device under investigation. The scale bar in (\textbf{a}) represents 500 nm. (\textbf{c}) and (\textbf{d}) schematically illustrate the band diagrams for the formation of the quantum dot in the electron-transport and hole-transport regimes, respectively.}
\end{figure}

A quantum dot is formed in the InSb nanowire segment sandwiched between the Nb contacts due to the combined interplay of native charges, metal depletion, and screening effects. As shown in Fig.~\ref{Fig1}(c), applying a positively biased gate voltage creates a potential well profile, specifically an $n$-type quantum dot. Similarly, applying a negative gate voltage can result in the formation of a $p$-type quantum dot. Electrons residing in the leads can tunnel into and out of this $p$-type quantum dot through the Zener-Esaki tunneling mechanism, as depicted in Fig.~\ref{Fig1}(d).

\begin{figure}
\centering
\includegraphics[width=8.5cm]{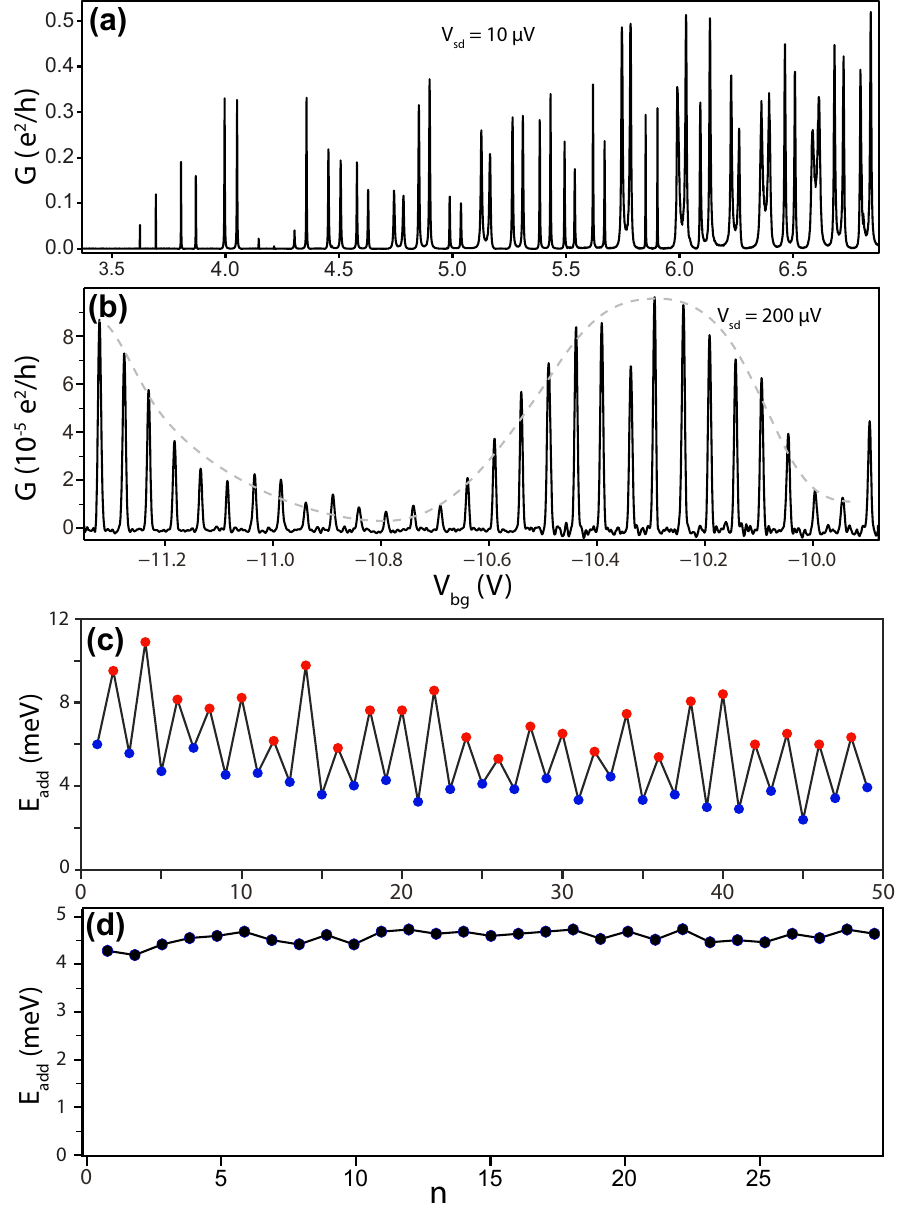}
\caption{\label{Fig2} \bt{Coulomb oscillations in $n$-type and $p$-type regimes.} (\bt{a}) Linear-response conductance $G$ measured for the device shown in Fig.~\ref{Fig1} at a positively-biased back gate voltage regime, with $V_{\rm sd}$ fixed at 10~$\mu$V. The conductance peaks are attributed to Coulomb oscillations in the electron-transport regime. (\bt{b}) Same with (\bt{a}), but measured at a more negative back gate range, with $V_{\rm sd}$ fixed at 200~$\mu$V. The conductance peaks here can be ascribed to Coulomb oscillations in the hole-transport regime. The oscillation peak envelope (dashed-line) can be attributed to the shifting of the lead-states in the nanowire, see the main text below. (\bt{c}) Addition energy extracted from Coulomb oscillations in panel (\bt{a}) as a function of the number of electrons in the quantum dot. An even-odd alternation behavior of the addition energy is observed. (\bt{d}) Addition energy extracted for Coulomb oscillations in panel (\bt{b}). No even-odd alternation is visible in the hole-transport regime.}
\end{figure}

We initially characterize the device by measuring the current at a low source-drain voltage, $V_{\rm sd}$, while varying the back gate voltage, $V_{\rm bg}$. When a quantum dot level aligns with the Fermi level of the leads, it gives rise to a distinctive peak in resonant tunneling conductance. As shown in Figs.~\ref{Fig2}(a)-(b), around 50 peaks are observed in the $n$-type regime, whereas 29 peaks emerge in the $p$-type regime. Notably, due to carriers having to tunnel through the inter-band gap in the $p$-type configuration, there is a significantly lower tunneling current compared to that of the $n$-type regime even when applying a larger bias voltage $V_{\rm sd}$.

\begin{figure*}
\centering
\includegraphics[width=18cm]{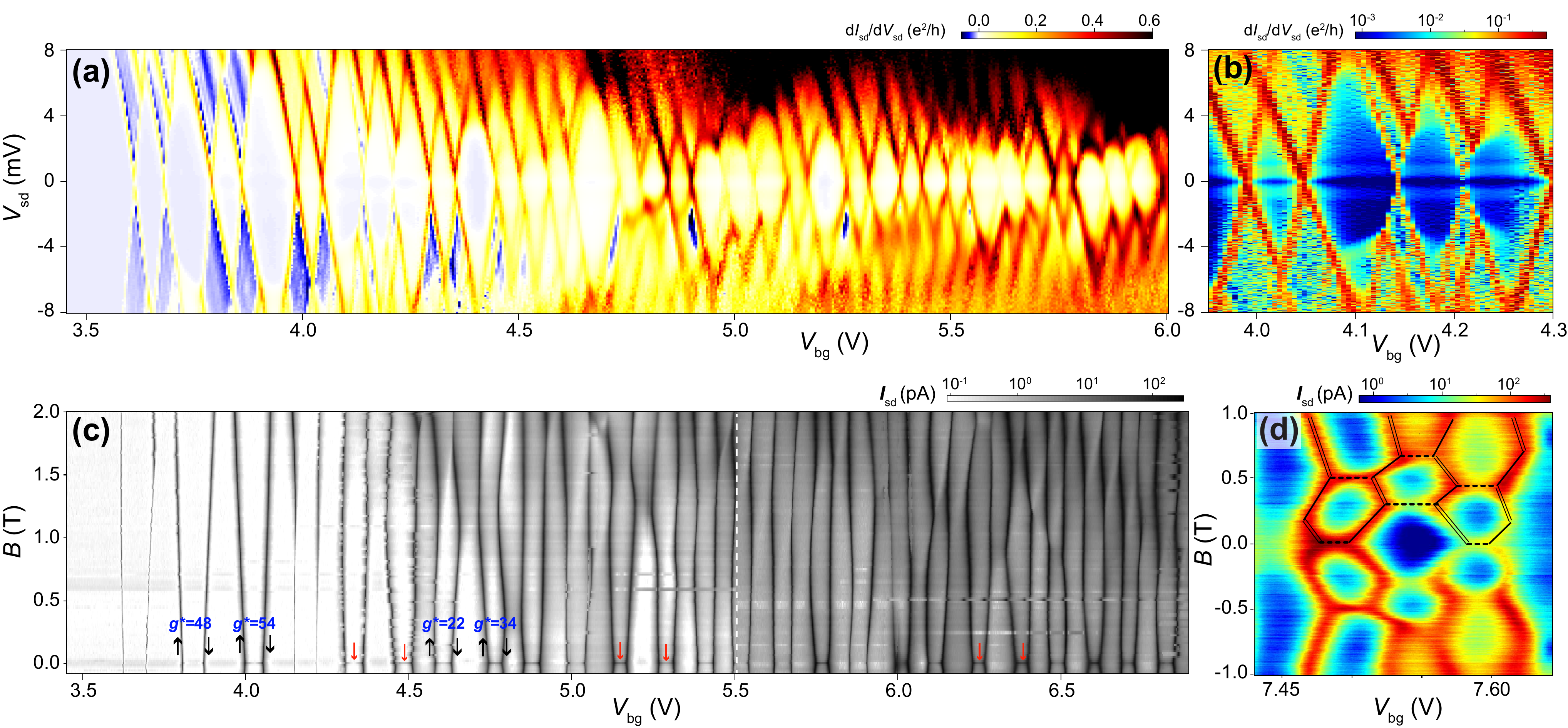}
\caption{\label{Fig3} \bt{Characterization of the electron-transport regime}. (\bt{a})  Charge stability diagrams in the electron-transport regime. Successive Coulomb diamond structures are clearly shown. The spectrum from $V_{\rm bg}=4.5\sim6.0$V is the same with Fig.5 in the supplementary material of Ref.~\onlinecite{Deng2014}.  (\bt{b}) A close-up view of (\bt{a}) in log-scale, to illustrate the superconductivity-induced conductance peaks. (\bt{c}) Quantum level evolution in magnetic field, corresponding to Fig.~\ref{Fig2}(\bt{a}). Here, the Zeeman effect induced level-shifting is evident. Blue numbers are a few extracted $g$-factors. At zero magnetic field, Kondo-effect induced conductance enhancement in the odd-occupied regimes can also be seen. (\bt{d}) Quantum level evolution in the magnetic field with multiple Kondo-peaks, where single-solid-lines indicate spin-down quantum levels, double-lines indicate spin-up quantum levels, and dashed-lines are Kondo-peaks. Note that panel (\bt{c}) comprises two individual measurements, stitched at $V_{\rm bg}=5.5$~V (indicated by the dashed line), to relieve the gate voltage hysteresis effect due to large-range sweep. To compensate the gate hysteresis, the high-$V_{\rm bg}$ part is offset to the left by 157~mV. } 
\end{figure*}

We can determine the addition energy of the quantum dot, denoted as $E_{\rm add}$, by analyzing the spacing between conductance peaks. Figs.~\ref{Fig2}(c)-(d) summarize the extracted $E_{\rm add}$ values as a function of the carrier filling number. Notably, in the $n$-type regime, the addition energy exhibits a distinct alternating pattern of large and small values. This alternating behavior is characteristic of a quantum dot with two-fold degenerate quantized levels, such as orbital levels with spin degeneracy.

For odd filling numbers, the corresponding $E_{\rm add}$ is primarily determined by the electron charging energy, denoted as $E_c^n$ (where the superscript n indicates the electron-transport regime). Conversely, for even-occupation quantum dots, the $E_{\rm add}$ comprises both the $E_c^n$ term and the quantum energy difference between adjacent orbital levels, $\Delta\epsilon$. Based on this analysis, we can extract the $E_c^n$ term to be in the range of 2.3$\sim$6 meV and the $\Delta\epsilon$ to be 1.2$\sim$6.1 meV.

In contrast, the hole-transport regime does not exhibit any visible even-odd alternation, indicating that $E_{\rm add}$ is predominantly influenced by the hole charging energy, $E_c^h$, in this regime. We can extract $E_c^h\approx 4.2\sim4.7$meV from Fig.~\ref{Fig2}(d). Since the effective mass of InSb heavy holes in the valence band is significantly larger than that of conductance electrons and light holes, it suggests that the quantum dot is predominantly populated by heavy holes due to the small energy quantization~\cite{Vurgaftman2001}.

We now focus on the electron transport regime. As illustrated in Figure~\ref{Fig3}(a), the charge stability diagram, depicting the dependence of differential conductance $dI_{\rm sd}/dV_{\rm sd}$ on $V_{\rm sd}$ and $V_{\rm bg}$, reveals a distinct pattern characterized by diamond-shaped structures. According to the ratio of the expansions in $V_{\rm sd}$ and $V_{\rm bg}$ within each individual Coulomb diamond, we estimate the lever arm of the back gate to be approximately $\alpha_{e} \sim 0.1$. Furthermore, in the low $V_{\rm bg}$ region, there are prominent negative differential conductance strips (highlighted in blue), which will be addressed later.

Note that, owing to the superconducting nature of the Nb-based contacts, a low-conductance strip is observed in the low-$V_{\rm sd}$ region, as evident in Fig.~\ref{Fig3}(b). This low-conductance strip arises from the proximity effect, which induces a superconducting gap in the InSb nanowire segments located beneath the Nb-based electrodes. For a more comprehensive understanding of measurements within the superconductor gap for the device, see the supplementary material of Ref.~\onlinecite{Deng2014}.

\begin{figure*}
\centering
\includegraphics[width=18cm]{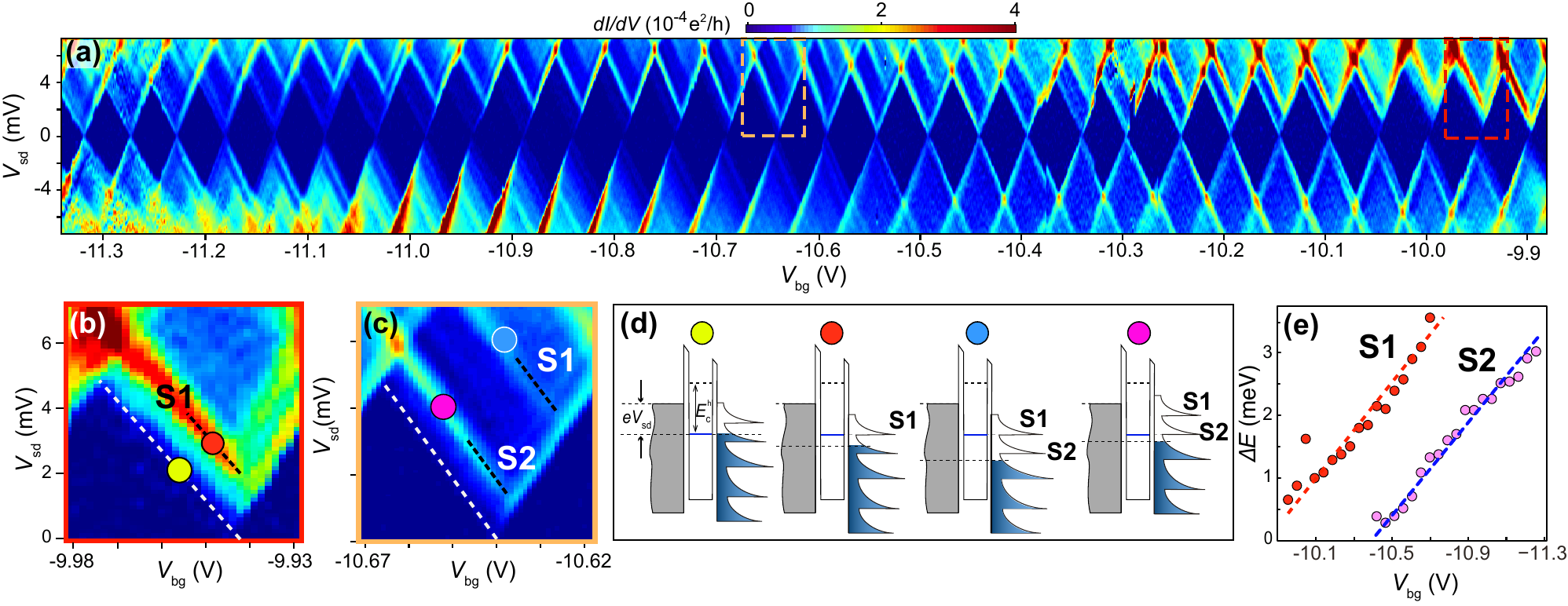}
\caption{\label{Fig4} \bt{Characterization of the hole-transport regime}. (\bt{a}) Charge stability diagrams in the hole-transport regime, where 30 regular Coulomb-diamond structures are shown. (\bt{b}) and (\bt{c}) Close-up view of the dashed rectangles in panel (\bt{a}) with corresponding colors, where Lead-state induced conductance peaks can be identified (black dashed-lines). (\bt{d}) Schematics represent the quantum dot configurations at the corresponding positions indicated by circles with colors in the transport diagrams. (\bt{e}) Lead-state energy (relative to the Fermi level of the lead) as a function of $V_{\rm bg}$, extracted by tracing lead-state induced peaks. Note there could be lead-states in both the left and the right leads, but we only picture and analyze one side for simplicity without loss of generality.}
\end{figure*}

The device is then characterized in a perpendicularly applied magnetic field (relative to the substrate). As depicted in Fig.~\ref{Fig3}(c), notable shifts in the quantum dot levels of the $n$-type regime are observed. When subjected to a weak magnetic field, the evolution of these levels is primarily governed by the Zeeman effect. Specifically, quantum levels associated with up (down) spins tend to shift towards lower (higher) energy states. The magnitude of the Zeeman shift is determined by the formula $E_{\rm Z}=\frac{1}{2}g^*\mu_B B$, where $g^*$ represents the Lande $g$-factor, $\mu_B$ denotes the Bohr magneton, and $B$ is the strength of the magnetic field. In Fig.~\ref{Fig3}(c), we have marked several spins (indicated by black arrows) and subsequently extracted the corresponding $g$-factors. These extracted values are notably large and vary depending on the level, aligning with previous findings reported in Ref.~\onlinecite{Nilsson2009}. However, as the magnetic field intensity increases, the evolution of these levels becomes more intricate due to the influence of the level interaction.  

Another observation in the evolution of the magnetic field is the appearance of high-conductance ridges when the filling number of the quantum dot is odd, which aligns with signatures of the Kondo effect~\cite{Goldhaber-Gordon1998, VanderWiel2000, Kretinin2011, Kanai2011}. For instance, in Fig.~\ref{Fig3}(c), several high-conductance ridges are highlighted by red arrows at zero magnetic field. As the magnetic field increases, these ridges progressively vanish, consistent with the suppression of Kondo correlations due to the lifting of spin degeneracy by the Zeeman effect. Similar high-conductance ridges have been observed in other InAs and InSb nanowire quantum dot devices~\cite{Csonka2008, Nilsson2009, Nilsson2010, Pillet2013, Deng2014, Deng2024}. Notably, in a comparable InSb quantum dot device reported in Ref.~\onlinecite{Deng2024}, temperature-dependent measurements confirmed that these high-conductance ridges are attributed to the Kondo effect.

Nevertheless, it is conceivable that Kondo correlations may re-emerge at higher magnetic fields when two different orbital levels with opposite spins align. We also observed the re-emergence of the high-conductance ridges in Fig.~\ref{Fig3}(d), where four ridges (indicated by dashed lines) become apparent at finite magnetic fields (at higher $V_{\rm bg}$ values). The ridges, along with the resonant tunneling peaks, form a ``honeycomb'' pattern. The ridges are highly likely re-emerged Kondo peaks in the magnetic field~\cite{Kanai2011}.


Now, we turn to the $p$-type regime. To further investigate the hole-transport region, we have measured the charge stability diagram and presented it in Fig.~\ref{Fig4}(a), corresponding to the back gate range in Fig.~\ref{Fig2} (b). Remarkably, nearly 30 continuous and stable Coulomb diamond structures are achieved without any charge rearrangement or impurity-induced interrupt. The small orbital energy difference $\Delta\epsilon$ leads to almost constant diamond sizes (in terms of both the $V_{\rm sd}$-extension and the $V_{\rm bg}$-extension). In this regime, the charging energy, $E_c^h$, is approximately 4.5~meV on average, with a lever arm factor of $\alpha_h\sim$0.09.

We have also observed high conductance peaks outside each Coulomb diamond, e.g., the peak marked by the red circle in Fig.~\ref{Fig4}(b). Conductance peaks outside Coulomb diamonds have been reported, which can be attributed to the tunneling process through excited states of the quantum dot~\cite{DeFranceschi2001}. However, the peaks outside Coulomb diamonds in Fig.~\ref{Fig4}(b) can not be ascribed to excited states because the small $\Delta\epsilon$ of the $p$-type regime does not match the energy difference between the ground state and the excited state. Instead, by tracking their positions relative to the Coulomb diamond edge, we attribute the out-diamond peaks to the electron density of states (DOS) in the leads, i.e., the lead-states or the reservoir-states~\cite{Mottonen2010, Bjork2004}. 

As shown by the energy diagrams in Fig.~\ref{Fig4}(b), the actual leads are the InSb nanowire segments under the Nb-based contacts [Fig.~\ref{Fig1}(c)-(d)]. Within the nanowire, the continuum approximation holds true only for the longitudinal direction, rendering its DOS quasi-one-dimensional (Q1D) with highly nonuniform characteristics. In the tunneling regime of the quantum dot, the current is directly proportional to the DOS of the reservoir at a given energy. The differential conductance can be significantly suppressed even at the edge of Coulomb blockade diamonds when the quantum dot level is aligned with the valley of the Q1D DOS (labeled with yellow circle), while peaks in differential conductance emerge when the quantum dot level aligns with spikes in Q1D DOS (red circle).

The Q1D lead-states can also be tuned by the back gate, although the lever arm is much smaller than the dot-level lever arm due to the screening effect of the metallic contacts. The lead-states are therefore trackable in the successive Coulomb diamonds. Conductance peaks originating from the same lead-state can emerge with different Coulomb diamonds, with a small energy shifting $\Delta E$ caused by $V_{\rm bg}$. The quantum dot can thus serve as a lead-state probe, and tomographically picture the dynamics of lead-states. The energy shiftings for two lead-states are extracted and shown in Fig.~\ref{Fig4}(e). The lead-state shifting also lead to an envelope for the Coulomb oscillation peaks in Fig.~\ref{Fig2}(b). We can calculate the back gate lever arm for the lead-states to be about 0.005, which is about 5\% of the lever arm for the dot states. 

The Q1D lead-state picture in the $p$-type regime is also in line with the negative differential conductance observed in the $n$-type regime [Fig.~\ref{Fig3}(a)]. The sharp lead-state DOS profile can modifies the tunneling current significantly and gives rise to the negative differential conductance~\cite{Mottonen2010}.

In summary, we have fabricated and measured an InSb nanowire quantum dot device with both $n$-type and $p$-type polarities. Using a back gate, we can tune the InSb nanowire device from the electron-populated quantum dot regime to the hole-populated quantum dot regime. Due to the high quality of the grown material and the clean fabrication process, tens of successive quantum levels are observed in both regimes without any charge rearrangement or impurity-induced interrupt. Coulomb blockade effect, Zeeman effect, Kondo effect, and lead-states are carefully studied. We have also established a tomographic characterization method for the lead-states by tracking the out-diamond peaks repeatedly appearing in the consecutive Coulomb diamond structures. This work is inspiring for the future application of ambipolar quantum devices.

\textbf{Acknowledgment.} This work was financially supported by the Swedish Research Council (VR), the Ministry of Science and Technology of China (MOST) through the National Key Research and Development Program of China (No.~2016YFA0300601 and 2017YFA0303304), the National Natural Science Foundation of China (Nos.~11874071, 91221202, 91421303, and 92165208), and the Innovation Program for Quantum Science and Technology (Grant No. 2021ZD0302401). 

\vspace{5pt}
M.T.D and C.L.Y contribute equally to this work.

\vspace{5pt}
\textbf{Contribution.} C.L.Y fabricated the devices. M.T.D and C.L.Y performed the measurements. P.C. grew the semiconductor nanowires. C.L.Y., M.T.D and H.Q.X analyzed the data. H.Q.X lead the experiments. All of the authors contributed to manuscript writing and discussion.

\appendix 

\section{Conductance spectra in magnetic fields}
\label{secAppendixA}

\begin{figure}
\centering
\includegraphics[width=8.5cm]{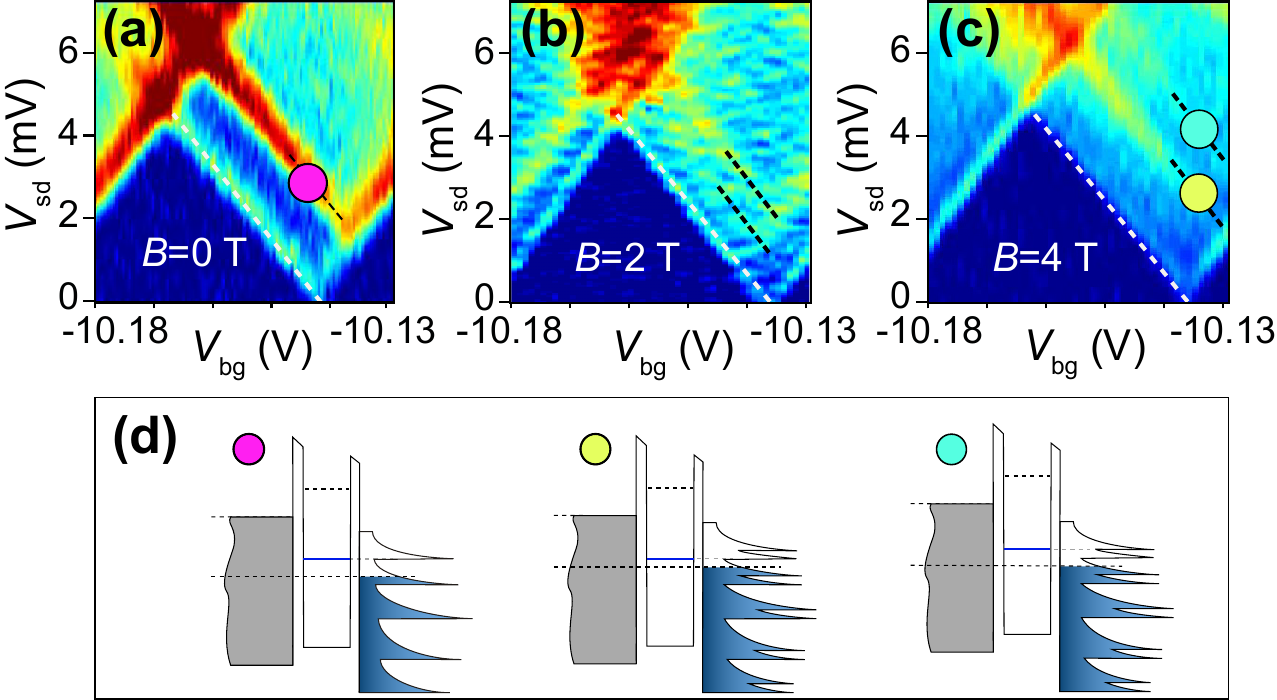}
\caption{\label{FigS1} \bt{Charge stability diagrams in the $p$-type regime of device-A. } (\bt{a-c}) Charge stability diagram measured at $B=0$T, $B=2$T and $B=4$T, respectively. The lead-state induced high-conductance line (purple circle) splits into two weak lines at $B=2$T and further splits at $B=4$T (yellow and cyan circles). (\bt{d}) Schematics of energy diagrams of the device.}
\end{figure}

We have also investigated the evolution of lead-state induced conductance peaks in response to magnetic fields. As shown in Fig.~\ref{FigS1}, the high conductance peak at $B=0$T (labeled with the purple circle) seemingly splits into two peaks at $B=2$T, and further splits at $B=4$T. Although the peaks are weak at high magnetic fields, they can be cautiously attributed to the Zeeman splitting of the leads (cyan and yellow circles). Accordingly, we deduce the $g$-factor for the lead-state to be $g\sim $9.5. The extracted lead-state $g$-factor is significantly smaller than the dot-state $g$-factor in Fig.~\ref{Fig3}(c). The small $g$-factor can be ascribed to the superconductor screening effect from the Nb-based contacts. Even with an underestimation, the $g$-factor is still higher than that of InSb holes~\cite{Pribiag2013}. This is consistent with our assumption that the lead states are attributed to electron states in Nb-covered InSb nanowire segments.

\section{Data from device-B}
\label{secAppendixB}

\begin{figure*}
\centering
\includegraphics[width=18cm]{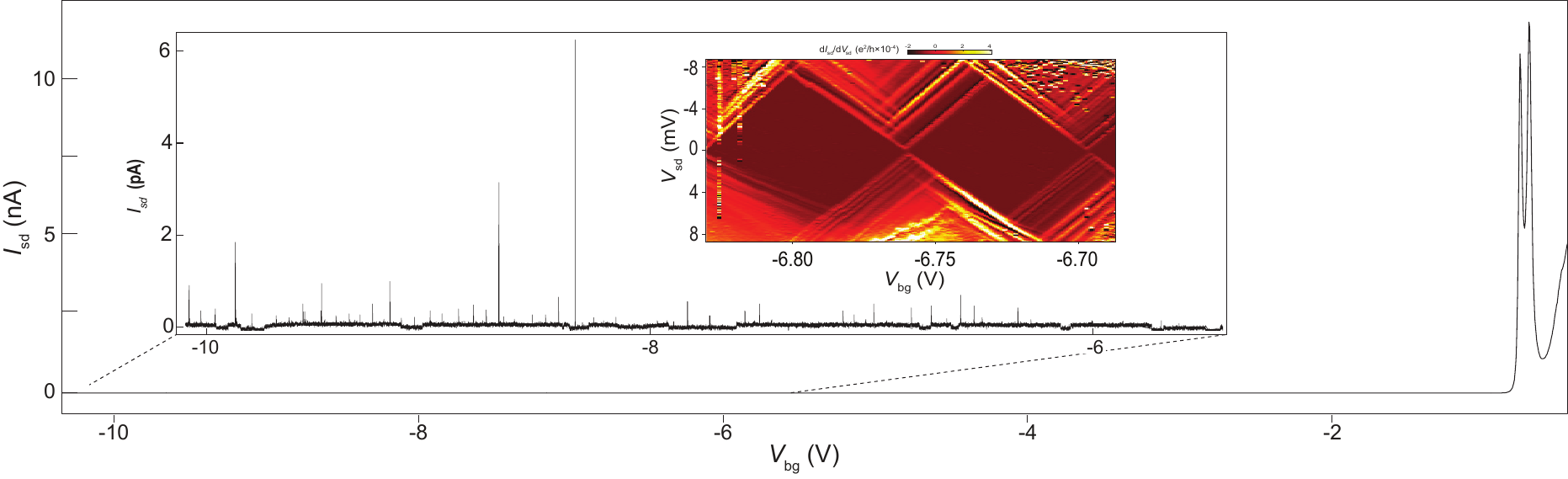}
\caption{\label{FigS2} \bt{Ambipolar quantum transport measurements from another InSb quantum dot device (device-B)}. Main plot: Current measured in the linear response region as a function of back gate voltage. Coulomb oscillation peaks from both the $n$-type regime (the high peaks in the right side) and the $p$-type regime (the small peaks in the left side and the Coulomb stability in the inset) can be identified. The $p$-type Coulomb peaks start to show up when $V_{\rm bg}<-6$V and are several orders of magnitude lower than the $n$-type peaks. Charge-stability diagrams for the $p$-type regime is shown in the inset.} 
\end{figure*}

Signatures of ambipolar quantum transport are also observed from another InSb quantum dot device, denoted as device-B. Device-B has a geometry similar to that of device-A, but with Al leads. Device-B in superconducting state has been reported in Ref.~\onlinecite{Deng2024}.

As shown in Fig.~\ref{FigS2}, device-B also possesses an ambipolar quantum transport nature. As the gate voltage increases from around 0V to $-10$V, the device undergoes a transition from a $n$-type quantum dot to a $p$-type quantum dot. No Coulomb peaks are observed in the range of $V_{\rm bg}=-0.8\sim-6$V. Note that the $p$-type Coulomb peaks are several orders of magnitude lower than the $n$-type peaks, similar to device-A. Lead-states are difficult to identify for device-B due to the short lead length and the irregular Coulomb diamond relative to device-A.

\bibliography{Bibfile}

\end{document}